\documentclass[aps,twocolumn,pre,superscriptaddress,showkeys,nofootinbib]{revtex4-2}
\usepackage{hyperref}
\hypersetup{colorlinks=true, linkcolor=blue, filecolor=magenta, urlcolor=blue, citecolor=blue}

\usepackage{graphicx}
\usepackage{dcolumn}
\usepackage{color} 
\usepackage{bm}
\usepackage{soul} 

\usepackage[mathlines]{lineno}

\begin{document}

\title{CuXASNet: Rapid and Accurate Prediction of Copper $L$-edge X-Ray Absorption Spectra Using Machine Learning}

\author{Samuel P. Gleason}
\affiliation{National Center for Electron Microscopy, Molecular Foundry, Lawrence Berkeley National Laboratory, Berkeley, CA, USA, 94720}
\affiliation{Department of Chemistry, University of California, Berkeley, CA, USA}

\author{Matthew R. Carbone}
\affiliation{Computing and Data Sciences Directorate, Brookhaven National Laboratory, Upton, NY 11973, USA}

\author{Deyu Lu}
\affiliation{Center for Functional Nanomaterials, Brookhaven National Laboratory, Upton, NY 11973, USA}

\author{Jim Ciston}
\affiliation{National Center for Electron Microscopy, Molecular Foundry, Lawrence Berkeley National Laboratory, Berkeley, CA, USA, 94720}

\date{\today}
\begin{abstract}
In this work, we have developed CuXASNet, a dense neural network that predicts simulated Cu $L$-edge X-ray absorption spectra (XAS) from atomic structures. Featurization of the Cu local environment is performed using a component of M3GNet, a graph neural network developed for predicting the potential energy surface. CuXASNet is trained on simulated spectra from FEFF9 at the multiple scattering level of theory, and can predict the $L_3$ and $L_2$ edges for Cu sites to quantitative accuracy. To validate our approach, we compare 14 experimental spectra extracted from the literature with the predictions of CuXASNet. The agreement of CuXASNet with experiments is shown by an average MAE of 0.125 and an average Spearman's correlation coefficient of 0.891, which is comparable to FEFF9's values of 0.131 and 0.898 for the same metrics. As such, CuXASNet can rapidly generate a large number of $L$-edge XAS spectra at the same accuracy as FEFF9 simulations. This can be used as a drop-in replacement for multiple scattering codes for fast screening of candidate atomic structure models of a measured system. This model establishes a general framework for Cu XAS prediction, and can be extended to more computationally expensive levels of theory and to other transition metal $L$-edges.
\end{abstract}
\maketitle


\section{Introduction}

X-ray absorption spectroscopy (XAS) is commonly used to identify atomic and electronic information about functional materials, including local chemical environments and oxidation states (OS)~\cite{DeGroot2008CoreSolids}. In XAS, a core-level photoelectron is excited to the conduction band. The spectrum serves as a fingerprint for the absorbing element's atomic number and is sensitive to core-level orbital energy changes caused by differences in the bonding environment~\cite{DeGroot_1990_2p}. The information contained in an XAS spectrum is crucial to the mechanistic understanding and rational design of functional materials, such as catalysts~\cite{Bai_2022, Kubin_2018}, renewable energy materials~\cite{Marelli2024Reaction, Pacheta2023EvolutionFilms} and batteries~\cite{Jenkins2023DirectBatteries}. XAS analysis is typically conducted by dividing the spectrum into two components. The first component is the x-ray absorption near-edge structure (XANES), and provides information on OS and the coordination environment of the absorbing site~\cite{DeGroot2008CoreSolids}. The second component is the extended X-ray absorption fine structure (EXAFS), and provides information on atomic structure beyond the absorbing site, such as bonding structure and nearest neighbor bond lengths~\cite{Yano_2009_X_ray}. This work focuses primarily on the generation and analysis of spectra for XANES studies.   

\begin{figure*}[ht]
    \centering
    \includegraphics[width=\textwidth]{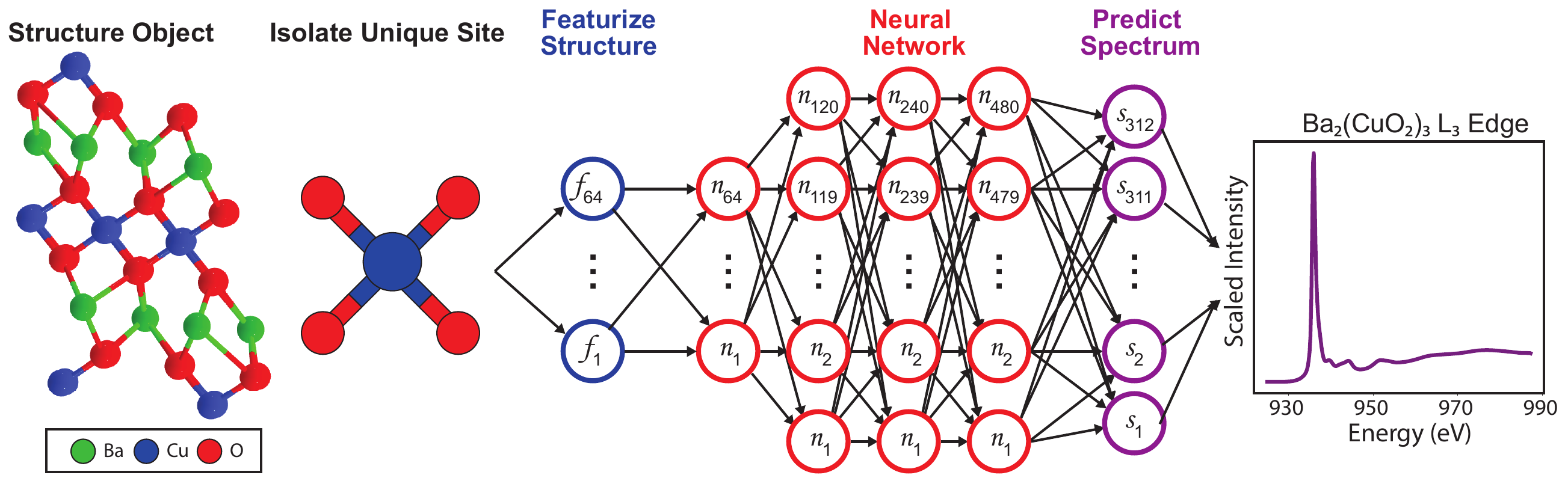}
    \caption{Outline of the CuXASNet model architecture. A structure object of a Cu containing material from the Materials Project (mpid: mp-615789) is represented using Pymatgen~\cite{Ong_2015_Materials}. Then, every symmetrically unique Cu site is extracted using the space group symmetry. Upon isolation of the unique Cu sites, they are featurized using M3GNet which transforms the site into a 64-dimension vector representing its local environment~\cite{Kharel_2024_Universal}. This vector is then input to our spectral prediction neural network, which contains 3 fully connected hidden layers, before returning a predicted spectrum that is discretized on an energy grid of 312 dimensions.}
    \label{fig:model_outline}
\end{figure*}

Many technologically important functional materials contain $3d$ transition metals, such as Ti, Mn, Fe, Co, Ni and Cu. This is primarily due to their chemical flexibility, namely that they can adopt several different OS. Additionally, these elements are relatively highly abundant, allowing for simpler commercial development relative to less abundant elements, such as Pd, Rh, and Ru~\cite{wheelhouse_2023_advances}. $3d$ transition metals are common in catalytic materials, which often contain $3d$ transition metal centers.  For instance, Ni, Mn and Fe are used in artificial photosynthesis~\cite{Dalle_2019}, Cu is used in in CO$_2$ reduction~\cite{Gao_2023_Experimental} and the azide-alkene cycloaddition click reaction~\cite{Gawande_2016,Dervaux_2012_Heterogeneous,Astruc_2012_Click}, and Cu, Fe and Ti are used in photovoltaic devices~\cite{Mccusker_2019,Liu_2018_Recent, Frye_2023_Reaching,Regan_1991_Low}. $3d$ transition metals are also common in biotechnology and medical applications. For example, Co has attained wide usage in medical imaging~\cite{Renfrew_2017_Harnessing}, Cu is present in many antimicrobial devices~\cite{Graham_2024_Towards}, and molecules with Fe centers are studied for a wide range of biomedical applications, including drug delivery and cancer treatment~\cite{Wang_2021_Novel,Cui_2021_Outstanding}.     

\begin{figure*}[ht]
    \centering
    \includegraphics[width=\textwidth]{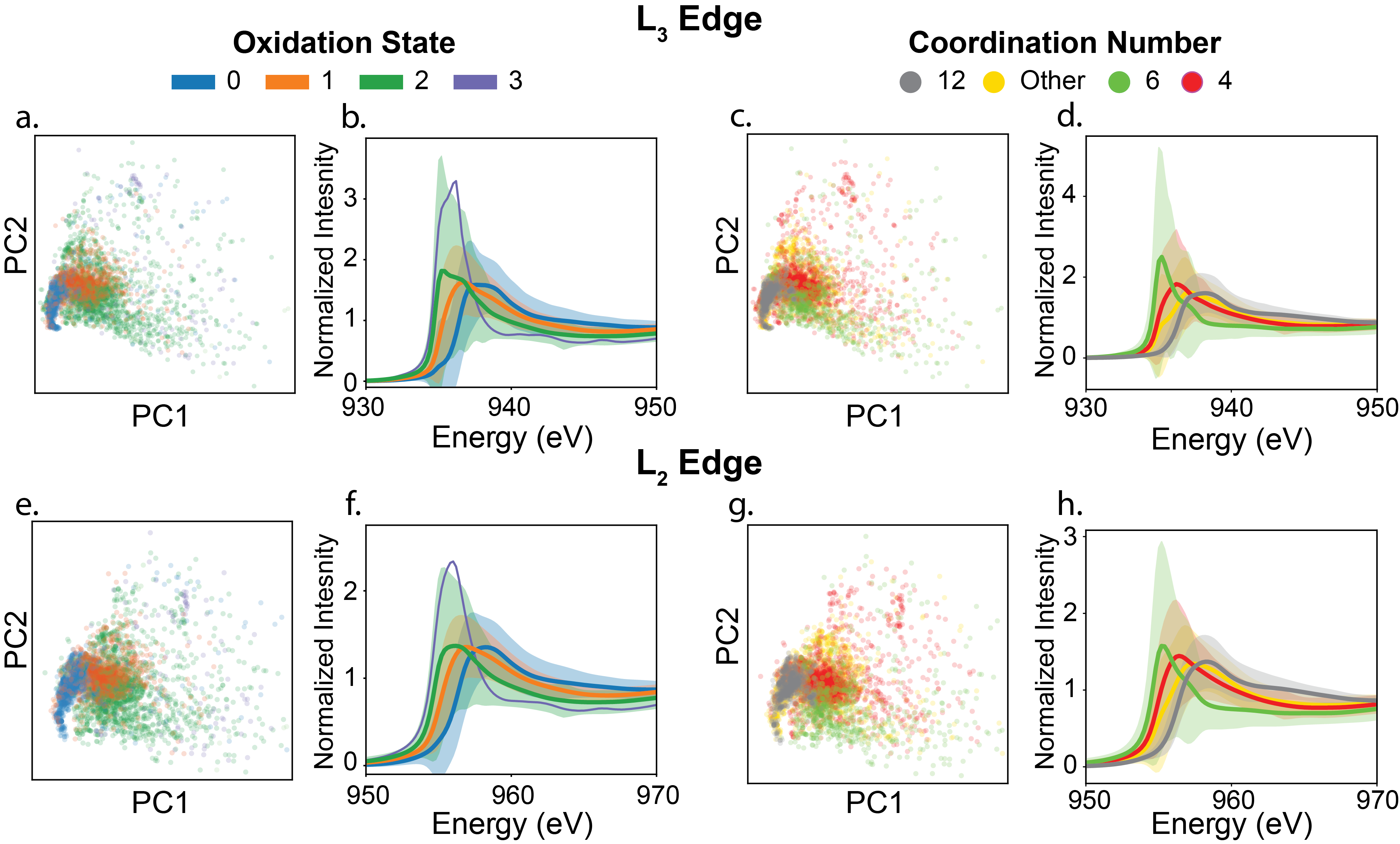}
    \caption{Illustration of the data spread in our spectral dataset. Each row shows a principle component analysis (PCA) and spectral comparison of the $L_3$ (a-d) and $L_2$ (e-h) edges, colored by their OS and CN. The first and third columns show scatter plots of the OS (a and e) and coordination number (c and g) of the top two principle components in the PCA conducted on the $L_3$ and the $L_2$ edges. The second and fourth columns show the average spectrum (solid line) across each OS (b and f) and coordination number (d and h) for $L_3$ and $L_2$ edges and spectra within one standard deviation (shaded area around the solid line). 
    }
    \label{fig:dataset_visualization}
\end{figure*}

XAS analysis of $3d$ transition metals is typically done using the $K$-, $L$-, and $M$-edges, which correspond to the excitation of core electrons from the $1s$, $2p$, and $3d$ orbitals, respectively~\cite{DeGroot2008CoreSolids}. Among these, the $L$-edge and the $K$-edge are commonly used for structural analysis of $3d$ transition metals~\cite{Baker2017K-Sites}. $K$-edge spectra are commonly used for for biological and organic chemistry samples, due to the damage caused by soft X-rays used in $L$-edge excitation on organic and biological materials~\cite{Gianoncelli_2015_Soft}, which is a consequence of the high soft X-ray absorption cross section of C, N, and O. However, for most materials science applications, the analysis of $3d$ transition metals using XAS is often focused on the $L$-edge~\cite{Kubin_2018}. The creation of a core hole in a $2p$ orbital is more stable than the $1s$ orbital, thus resulting in significantly increased maximum energy resolution due to the decrease in lifetime broadening~\cite{Baker2017K-Sites}, enabling observation of additional details in the fine structure of spectra. Additionally, the high energies associated with the $3d$ transition metal $K$-edge, which ranges from roughly 5~keV for Sc to roughly 10~keV for Zn, result in worse energy resolution than the lower energy $L$-edge~\cite{Baker2017K-Sites}. The transition metal K-edge is also prohibitively high in energy for other core level spectroscopy measurements, such as electron energy loss spectroscopy (EELS), due to the inability of electron detectors to effectively measure this energy range.  

Analysis of $L$-edge XAS is typically done via matching simulated spectra or experimental standards of known chemical composition and atomic arrangement to an unknown sample. In the case of experimental spectra, a few online databases of experimental spectra exist~\cite{Cibin_2020_Open,IXAS}. However, these are often focused on a few specific systems and are too narrow in scope to be used for general experimental spectral analysis. Additionally, they often contain mainly $K$-edge data for $3d$ transition metals. The limitation of matching unknown experimental samples to simulated spectra is typically the accuracy of the simulations, which often varies depending on the atomic structure and the simulation methods~\cite{Chen_2021,Vinson_2011_Bethe,Vinson_2022_Advances,Zheng_2020_Random}. Additionally, the availability of simulated data, particularly for more computationally intensive simulations, is often a significant limitation. To acquire accurate simulations that are useful for experimental analysis, detailed knowledge of the material structure is required~\cite{Lightshow,FEFF9_source}. Due to the computational cost of simulations, performing extensive searches of candidate atomic structure models in the high-dimension configuration space that match with target measured XAS spectra, either through brute force methods or structure sampling algorithms, is computationally prohibitive. Therefore, there is a need for a spectral prediction model that can rapidly generate accurate simulated spectra from an arbitrary atomic structure. 

\begin{figure*}[ht]%
    \centering
    \includegraphics[width=\textwidth]{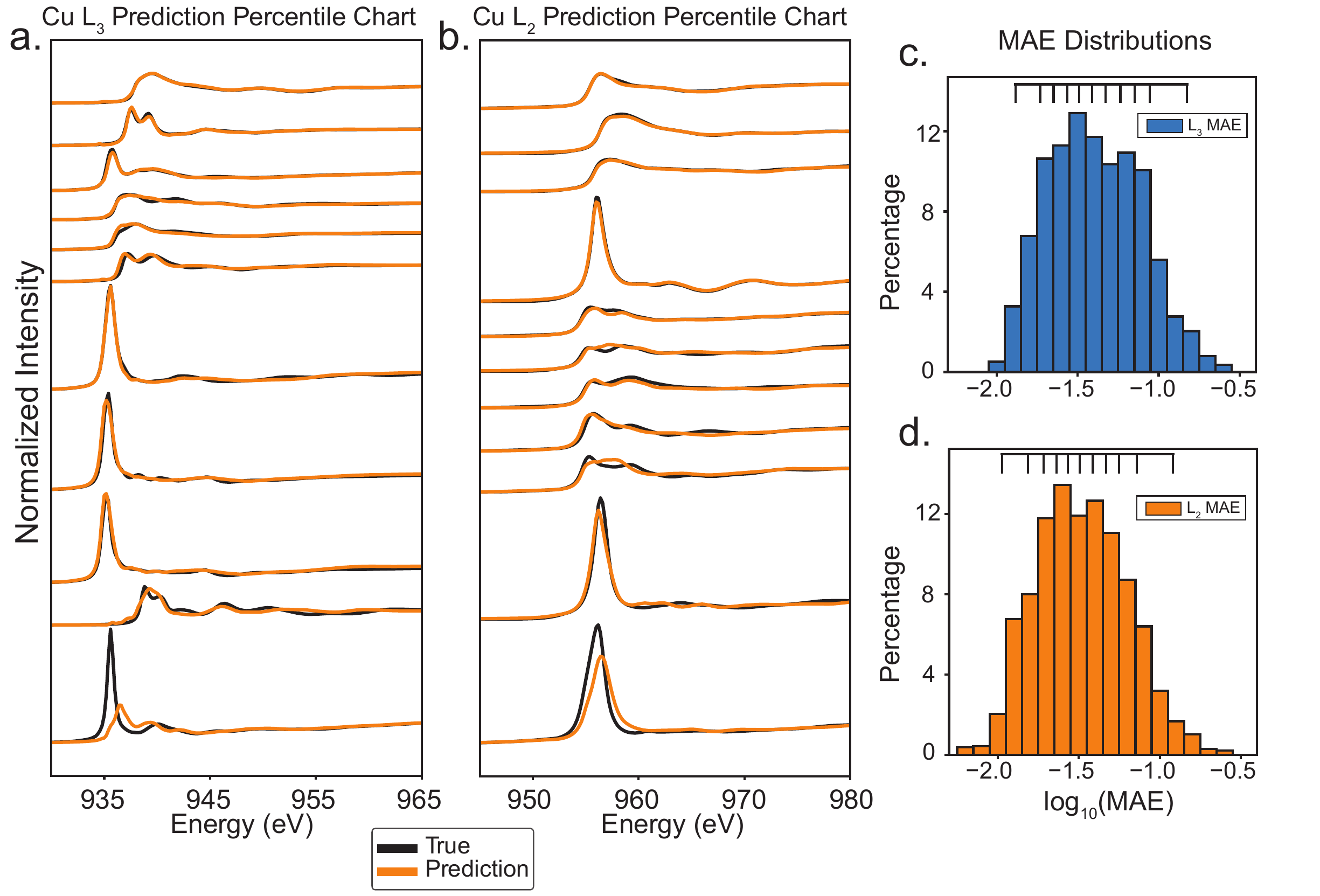}
    \caption{Waterfall plots showcasing the accuracy of our model at reproducing spectra simulated with FEFF9. (a-b) show decile charts with decreasing accuracy moving down the chart.  The spectra visualized in these plots reflect the middle 95\% of the test set, removing the best and worst 2.5\%. Each comparison shows a decile value of CuXASNet's performance on the middle 95\% of our test spectra, with the first comparison showing the most accurate included prediction, the second showing the 10th percentile of this subset of the test data, and subsequent spectra being spaced by 10 percentile increments. The $L_3$ and $L_2$ edges are shown in (a) and (b), respectively. The MAE distributions are shown in (c-d), which have been projected onto a log scale. The horizontal tick marks correspond to spectra in (a) and (b), with moving rightward across (c) corresponding to moving down in (a), and the same being true for (d) and (b).}
    \label{fig:sim_spec_accuracy}
\end{figure*}

A large collection of simulated XAS spectra is found in the Materials Project~\cite{Jain2013} Database, which includes $K$-edge~\cite{Zheng_2018} and more recently $L$-edge~\cite{Chen_2021} XAS spectra of a variety of different materials using the FEFF9 code~\cite{FEFF9_source,Mathew_2018,Zheng_2018,Jain2013}. However, despite the over $\sim$100k new $L$-edge spectra generated by \emph{Chen et al.}~\cite{Chen_2021}, the vast majority of structures in the Materials Project do not have associated $L$-edge spectra due to the computational cost of XAS simulations. For example, out of the nearly 10k Cu-containing materials in the Materials Project, roughly 8.5k do not have an $L$-edge spectrum~\cite{Jain2013}. Therefore, there is a strong demand for models which can generate simulated data quickly enough to process well over a million $3d$ transition metal-containing structures while retaining enough accuracy for application to experimental analysis. Additionally, certain spectral features in the $L$-edge spectra of Ti and Fe can be better reproduced by levels of theory other than the multiple scattering method, and further investigation of these systems can necessitate more computationally intensive simulations~\cite{Chen_2021,Vinson_2011_Bethe,Vinson_2022_Advances}. For example, the Ti $L$-edge of TiO$_2$ is more accurately generated by the OCEAN simulation code compared to FEFF9~\cite{Chen_2021, Vinson_2022_Advances}. However, the computational cost of OCEAN simulations increases drastically with increasing system size. In systems with large number of atoms (e.g., several hundred atoms) and magnetic ordering in the unit cell, the computational cost can grow very rapidly~\cite{Vinson_2011_Bethe,Gilmore_2015_efficient}. Therefore, generating large volumes of OCEAN spectra is even more challenging than building a FEFF9 database. 

Beyond direct matching of simulated spectra to experimental spectra with incomplete or unknown structural information, simulated spectral datasets can also be used to train machine learning (ML) models for automated analysis of core-level spectroscopy data. Due to the increasing availability of simulated data, ML has seen explosive growth in recent years as a broadly applicable technique to analyze XAS data~\cite{Mathew_2018,Zheng_2018,Jain2013}. Several groups have used ML to predict coordination environments and 3-dimensional molecular geometries from XAS $L$-edge and $K$-edge spectra~\cite{Torrisi_2020_random,Timoshenko_2017_Supervised,Zheng_2020_Random,Tetef_2021_Unsupervised,Tetef_2022_Informed,Higashi_2023_Extraction,Carbone_2024_Accurate}, extract OS information from XAS~\cite{Gleason_2024_Prediction,Chatzidakis_2019,del_Pozo_Bueno_2023}, and determine structural information, such as bond length and angle, from core-level spectroscopy~\cite{David_2023_Towards,Guda_2021_Search,Guda_2021_Understanding}. However, the success of these ML approaches requires a large training dataset comprised of examples across relevant feature spaces. There are examples of automated core-level spectral analysis models trained exclusively on experimental spectra, such as in Ref.~\onlinecite{MnEdgeNet}, which predicts the OS of Mn. However, datasets for automated analysis of core-level spectra are more often built on simulated spectra due to the significant challenge of collecting a large volume of labeled experimental data. Additionally, the large scale development of core-level spectroscopy models has been limited in part by the smaller amount of available simulated data for the transition metal $L$-edge relative to $K$-edge spectra~\cite{Chen_2021,Jain2013}. 

Recently, machine learning models have been used to generate simulated XAS spectra at a fraction of the computational cost of direct simulation~\cite{Chen_2024_Robust,Rankine_2020_ADeep,Rankine_2022_Accurate,Watson_2022_Beyond}. These studies have mostly focused on $K$-edge simulations of small molecules/atomic clusters. Therefore, in this work we have developed CuXASNet, an ML model which generates $L$-edge XAS spectra of Cu materials using atomic structure as input (see Figure \ref{fig:model_outline}). We envision that CuXASNet can be used to significantly enhance the volume of simulated $L$-edge data of Cu materials available to the community. CuXASNet is trained on spectra simulated using FEFF9, which utilizes the multiple scattering simulation method~\cite{FEFF9_source}. We note that the methodology we present here is extendable to more computationally intensive simulations, such as OCEAN or VASP~\cite{Vinson_2011_Bethe,Vinson_2022_Advances,Gilmore_2015_efficient,Hafner_2008_Ab_initio}, and to methods based on multiplet ligand-field theory~\cite{DeGroot2008CoreSolids, Haverkort_2023_Multiplet}. 

The inherent problem in simulating XAS is learning the structure-property relationship. This is a non-trivial task when dealing with materials structures, due to the high dimensionality of the structure, including local disorder and defects that can cause localized changes in the symmetry and coordination. Therefore, the first step in constructing such a model is to featurize the structure into an input vector that can be mapped to the output vector, in this case the XAS spectrum. This structure featurization step has been explored by several groups in recent years, via molecular graph~\cite{Carbone_2024_Accurate} and materials graph~\cite{Kwon_2023_Spectroscopy}, the atom-centered symmetry function (ACSF)~\cite{Rankine_2022_Accurate}, smooth overlap of atomic positions (SOAP)~\cite{Rankine_2020_ADeep}, local many-body tensor~\cite{Kwon_2023_Harnessing} and M3GNet~\cite{Chen_2022_Universal}. M3GNet was selected as the structure featurizer in this work due to recent results showing it was the most effective model at dimensionality reduction for predicting XAS spectra~\cite{Kharel_2024_Universal}.   

Beyond the utility of generating simulated reference databases and training data for ML models, the simulated spectra generated by CuXASNet can be used for theoretical studies of Cu $L$-edge core-level spectroscopy and for the rational design of materials. CuXASNet generates an XAS spectrum from a specified structure, which allows large-scale studies of the impact of atomic structural changes on Cu's core-level spectrum. This can be highly relevant in functional material development, especially in semiconductor engineering, where minor structural changes involving specific atom substitution with dopants and the introduction of atomic level defects are a primary method of engineering specific electronic properties. Additionally, through the use of a model that can generate thousands of spectra in minutes from atomic structures~\cite{Kwon_2023_Spectroscopy}, rational design of Cu-containing materials can be performed by conducting a bulk search of relevant structures when a material with a specific set of properties is desired. The core-level spectrum can elucidate many applicable properties of a functional material, and CuXASNet can be used to determine structural candidates for specific applications.   

\section{Methods}
\subsection{Training Set Generation}
Pymatgen was used to extract simulated FEFF9 $L$-edge XAS spectra of Cu materials from the Materials Project Database~\cite{Jain2013, Ong_2015_Materials}. Currently, the Materials Project contains $L$-edge Cu XAS spectra of 1533 materials, which encompass 2300 site spectra from unique Cu absorbers~\cite{Chen_2021,Jain2013}. In order to ensure enough training data for training CuXASNet, roughly 2000 additional structures were extracted from the Materials Project by selecting all Cu-containing materials predicted to be stable by density functional theory (DFT) or that had been labeled as experimentally synthesized~\cite{Jain2013}. By filtering for stable and/or experimentally synthesized materials, we aimed to generate a dataset that would be most relevant to experimental studies of Cu materials. Using the \emph{Lightshow} workflow~\cite{Lightshow,meng_2024_multicode} and the FEFF9 code~\cite{FEFF9_source}, we previously simulated an additional 3387 site specific $L$-edge spectra of Cu absorbers~\cite{Gleason_2024_Prediction}. The unique Cu site(s) in each structure were determined using the space group symmetry via Pymatgen. The dataset was then further processed before model training. These processing steps include interpolation to place every spectrum onto a uniform grid with 0.2 eV energy scale and scaling all spectra to have the same energy range, 925 -- 987.2 eV and 945 -- 1007.2 eV for $L_3$ and $L_2$, respectively. These steps followed the same dataset construction and spectral processing procedure conducted in Ref.~\onlinecite{Gleason_2024_Prediction} and resulted in the same dataset, up to a few minor changes described in later sections. 


\subsection{Simulated Spectral Processing}
The spectral dataset was validated to retain only simulated spectra from converged FEFF9 calculations. Filtering unconverged FEFF9 calculations resulted in the removal of 89 spectra from the full dataset. Additionally, all materials used to compare the output of the CuXASNet model to experimental spectra were removed from the training set to ensure the model was not biased by having seen these spectra in the training process. Consequently, these spectra are not in the test set examined in the Results and Discussion section. The resulting dataset used for model training and validation contains 5497 site-specific spectra. 

\begin{figure*}[ht]%
    \centering
    \includegraphics[width=\textwidth]{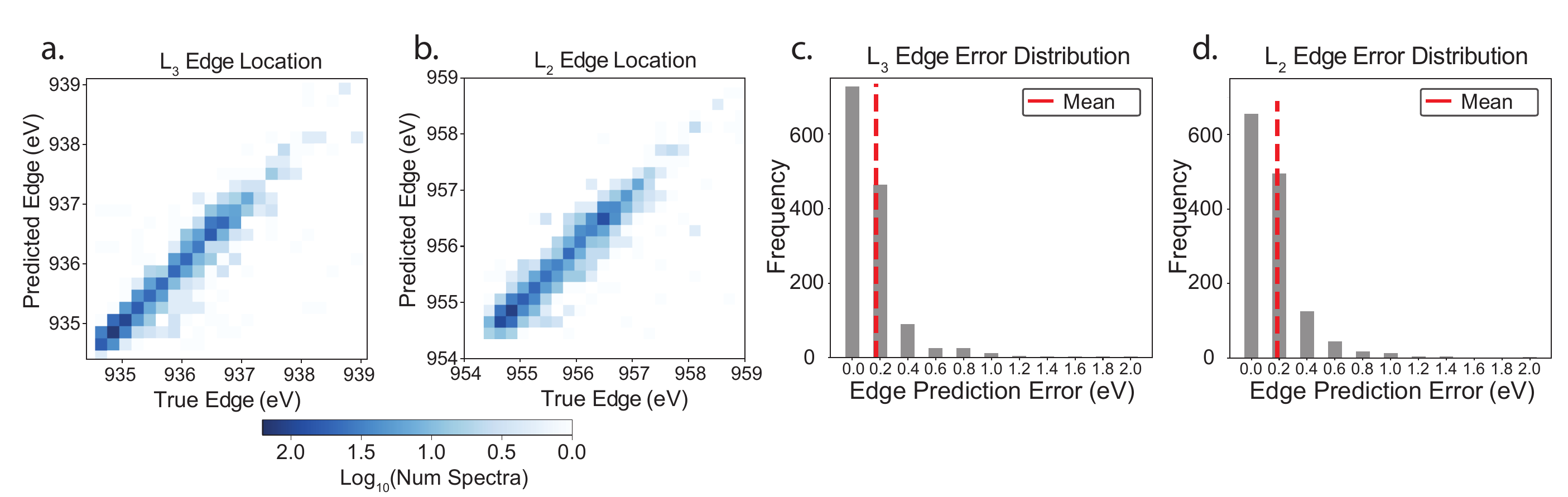}
    \caption{Accuracy of CuXASNet in predicting the FEFF9 simulated edge onset location of the $L_2$ and $L_3$ edge spectra, as determined by finding the location of the maximum in the first derivative in the first edge onset. 2-dimensional histograms of predicted versus true edge location are shown in (a-b), where the density is on a log scale to allow visualization of lower accuracy regions. Bar plots of the error in this prediction are shown in (c-d) on a 0.2~eV scale corresponding to the energy resolution of our simulated spectra.}
    \label{fig:edge_accuracy}
\end{figure*}

CuXASNet was trained using input vectors representing unique absorbing sites in an arbitrary structure object generated by the featurization module of M3GNet~\cite{Kharel_2024_Universal}. This was performed by extracting node-level features from the graph representation of a material's unit cell before M3GNet's readout layer (see Ref.~\onlinecite{Kharel_2024_Universal} for details). For each unique site, a feature vector of length 64, the default hidden state size for M3GNet, was generated to represent that specific site. This feature became the input vector for our ML model.  

\subsection{Dense Neural Network}
In CuXASNet, two deep neural network models were used to predict the $L$-edge XAS spectra of Cu absorbers from their featurized atomic structure, one for the $L_3$ edge and one for the $L_2$ edge with the same hyperparameters. Each neural network in CuXASNet has an input layer of length 64, equal to the length of the featurized structure. This is followed by three hidden layers of length 120, 240 and 480. The rectified linear unit (ReLU) activation function was used throughout, except for the output layers, where softmax was used. The output layer has a length of 312, with each point representing an intensity point in the spectrum separated by 0.2~eV. Models were trained using the Adam optimizer and mean squared error loss function for 500 epochs with a batch size of 10 in the training~\cite{Kingman_2014_Adam}. The number of epochs and batch size was not found to impact the model results. Batch sizes from 10-1000 did not change the median MAE of the test set, 0.0391, by more than 2\%, and a batch size of 10 was found to produce the best model. The number of epocs had a similarly small impact, with 500 producing the best model. igure~\ref{fig:training_loss} shows the relation of the training loss (between the predicted and FEFF9 spectra) versus epoch, where the model has converged by the end of the training process for both $L_3$ and $L_2$ edges. 

\subsection{Experimental XAS Data}
Experimental XAS from the literature were used to determine the applicability of the predictions from CuXASNet to experimental analysis. 14 experimental spectra were extracted from the literature by using webplotdigitizer~\cite{Marin2017, Goh_2006_oxidation, Liu_2001_Evidence, Buckley_2009_Electronic, Glaser_2018_Doping, Rudyk_2011_Electronic, Blanchard_2010_Electronic, Grioni_1992_Unoccupied}. The experimental spectra extracted for verification are the following: Cu metal, Cu$_2$O, CuO, Cu$_2$S, CuS, CoCu$_2$S$_4$, CuBe, CuFeS$_2$, LaCuOS, LaCuOSe, LaCuOTe, SrCuO$_2$, ZrCuSiAs and ZrCuSiP. The experimental spectra were processed using a broadening function that ensured the extraction produced a smoothed spectrum. Our simulated $L_{2,3}$ spectra are set on an energy range from 925 eV to 1007.2 eV, however, in multiple cases the experimental spectra were measured on a smaller energy range (i.e. 930 -- 975 eV). When this occurred, the simulated spectrum was compared to the experimental spectrum only over the range where experimental data was recorded. Then the simulated spectra were cropped to the same energy range and the mean absolute error (MAE) was calculated between the simulated/CuXASNet spectrum and the experimental spectrum.  

\section{Results and Discussion}
\subsection{Spectral Database Analysis}
The spectral database used in this work is comprised of 5497 site-specific Cu $L_{2,3}$-edge XAS spectra of across roughly 3400 Cu containing materials. The spectral features are illustrated in Figure~\ref{fig:dataset_visualization} by first distilling spectral features to a series of component vectors via principle component analysis (PCA) overlaid with averaged spectra across two Cu chemical descriptors, OS and coordination number (CN). PCA linearly decomposes a dataset into a set of principal vectors (components) and weights such that any spectrum can be reconstructed via a linear combination of the principal vectors. Plotting the weights of the most significant characteristic vectors for each spectrum and labeling them by known chemical descriptors can elucidate the degree of similarity across chemical systems, such as spectra that share the same OS. A detailed explanation of how PCA can be applied to spectral data can be found in Refs.~\onlinecite{Carbone_2019_Classification, Carbone_2024_Accurate}.

\begin{figure*}[ht]%
    \centering
    \includegraphics[width=\textwidth]{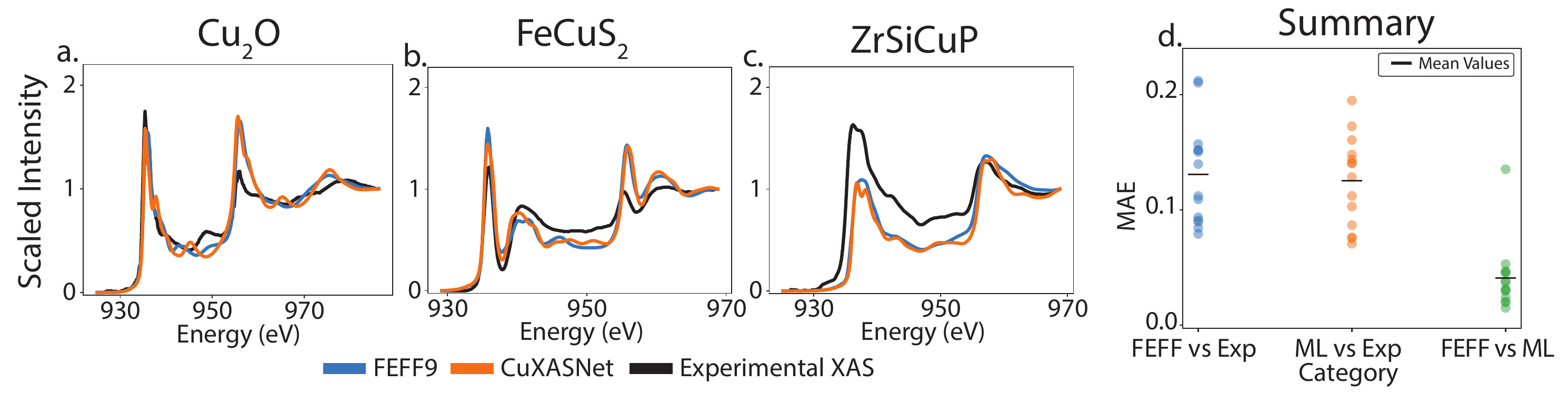}
    \caption{Performance of CuXASNet when compared to experimental spectra. Comparisons between FEFF9 simulations (blue), CuXASNet predictions (orange), and experimental spectra (black) are shown in (a) Cu$_2$O, (b) FeCuS$_2$ and (c) and ZrSiCuP. These three materials were chosen according to the best, a middle and the worst comparison of CuXASNet to experiments across our literature spectra. (d) Distributions of quantitative comparisons between FEFF9 and experiments (blue), ML and experiments (orange), and FEFF9 and ML (green).}
    \label{fig:experimental_comparison}
\end{figure*}

In Figure~\ref{fig:dataset_visualization}(a), the values of the first two principle components are plotted for each $L_3$ edge and colored by the OS of their corresponding absorbing site. The general trend shows that metallic materials, Cu(0), are highly clustered, indicating that spectral features of metallic Cu absorbers are relatively similar. As the oxidization state increases, this variability increases, although there are very few Cu(III) materials in this dataset due to the rarity of Cu$^{3+}$ ions in nature. Figure~\ref{fig:dataset_visualization}(b) shows the average of all spectra with a particular OS, with the spread of Cu(III) materials omitted due to their low frequency in the dataset. Figure~\ref{fig:dataset_visualization}(b) demonstrates the increased variance in Cu(II) as compared to Cu(I)/Cu(0), as shown by the larger standard deviation of the averaged Cu(II) spectrum. Figure~\ref{fig:dataset_visualization}(c-d) show a similar analysis with color coding and labels corresponding to CN. Figure~\ref{fig:dataset_visualization}(c) shows the same PCA plot as Figure~\ref{fig:dataset_visualization}(a) and reveals the following trends: Cu sites with a CN of 12 are strongly associated with metallic Cu in a face-centered-cubic (FCC) lattice, and this is confirmed by the averaged spectrum in Figure~\ref{fig:dataset_visualization}(d). Sites with a CN of 6 (e.g., an octahedron motif) are also reasonably well contained, while absorbing sites with CN of 4 (e.g., a tetrahedron motif) produce a wide range of features. Interestingly, there is more variance in the spectra with CN=4 than the `other' category, which encompasses many CNs that are all present in insufficient numbers to be visualized directly in Figure~\ref{fig:dataset_visualization} (see Figure \ref{fig:data_distribution}). Figure~\ref{fig:dataset_visualization}(e-h) shows the same analysis as Figure~\ref{fig:dataset_visualization}(a-d), instead on the $L_2$ edge. The trends are essentially the same, with the only systematic difference being that the $L_2$ edge exhibits more uniformity in general, exemplified by the increased clustering in the first two principle components and the smoother averaged spectra. 

\subsection{$L_3$/$L_2$ Spectra Prediction}
CuXASNet is comprised of two separate models trained on the simulated data: one to predict the $L_3$ edge and one to predict the $L_2$ edge. The accuracy of each of these tasks is shown in Figure~\ref{fig:sim_spec_accuracy}. Figure~\ref{fig:sim_spec_accuracy}(a-b) shows a decile plot of the middle 95\% of the test set performers, discarding the best and worst 2.5\%. This range was chosen due to observations that the best and worst 2.5\% were significant outliers and did not represent the overall performance of the model. This can be seen by the very low density of the highest and lowest MAEs in Figure \ref{fig:sim_spec_accuracy}(c-d). Particularly, the least accurate 2.5\%, and more specifically the two worst spectra, comprising the worst 0.1\% of the test set, were observed to match the FEFF9 simulation very poorly. This is likely caused by the very uncommon features observed in these two FEFF9 spectra (e.g., the energy location of the edge onset is roughly 10 eV higher than most other simulated spectra), and more details on the worst outliers can be found in Figure~\ref{fig:worst_spec}. As can be seen from Figure~\ref{fig:sim_spec_accuracy}(a), the first 20\% of the test set is predicted with essentially perfect accuracy (first 3 predictions in Figure~\ref{fig:sim_spec_accuracy}(a) and the first three ticks in Figure \ref{fig:sim_spec_accuracy}(c)). In the next 70\%, the main spectral features are correctly predicted, with some regularization visible around the finer features and a few minor intensity mismatches (next 7 predictions in Figure~\ref{fig:sim_spec_accuracy}(a) and the next seven ticks in Figure~\ref{fig:sim_spec_accuracy}(c)). The worst $L_3$ prediction in the middle 95\% of the test set, shown at the bottom of Figure~\ref{fig:sim_spec_accuracy}(c), is a more notable failure of the model, where the edge location and edge onset peak shape deviates significantly from the FEFF9 simulation. However, it is interesting to note that the $L_2$ 97.5th percentile is far more accurate than the $L_3$, resulting in only a mild edge location and peak shape misidentification as shown in Figure~\ref{fig:sim_spec_accuracy}(b). Although the waterfall plots in Figure~\ref{fig:sim_spec_accuracy}(a-b) show different spectra, the error of the $L_3$ and $L_2$ prediction for the same absorber is highly correlated across our testing data, as shown in Figure~\ref{fig:edge_correlation}.

\subsection{Edge Location Accuracy}
The previous section demonstrates that CuXASNet can successfully predicts the $L_2$ and $L_3$ edge of a Cu site in a material. However, an important and perhaps more intuitive metric for the model's accuracy is the successful prediction of the edge onset of the spectrum. The location of the edge onset is determined by finding the location of the maximum in the first derivative in the first edge onset location in both the $L_3$ and the $L_2$ edges. Figure~\ref{fig:edge_accuracy}(a-b) show bar plots of the error in the edge location, with roughly half being predicted exactly correct for both $L_2$ and $L_3$ edges. As the spectra are set on a 0.2~eV scale, the errors naturally move in units of 0.2~eV, explaining the jumps in the location of error values across the bar plots. The distribution of the errors is very similar between the $L_3$ and $L_2$, with their means occurring at almost identical values. Figure~\ref{fig:edge_accuracy}(c-d) show two-dimensional histograms of the edge location accuracy, which show similar trends to Figure~\ref{fig:edge_accuracy}(a-b). Additionally, Figure~\ref{fig:edge_accuracy}(c-d) shows that around 300 spectra, out of the 1375 in the test set, are correctly predicted or estimated with only minor errors at between 934.8 and 935~eV for the $L_3$ edge and 954.8 and 955~eV for the $L_2$ edge. Errors greater than 0.2~eV are quite rare, and errors of over 1~eV are so uncommon such that they are almost impossible to see in Figure~\ref{fig:edge_accuracy}(c-d), despite projecting the density onto a log scale.

\begin{figure*}[ht]%
    \centering
    \includegraphics[width=\textwidth]{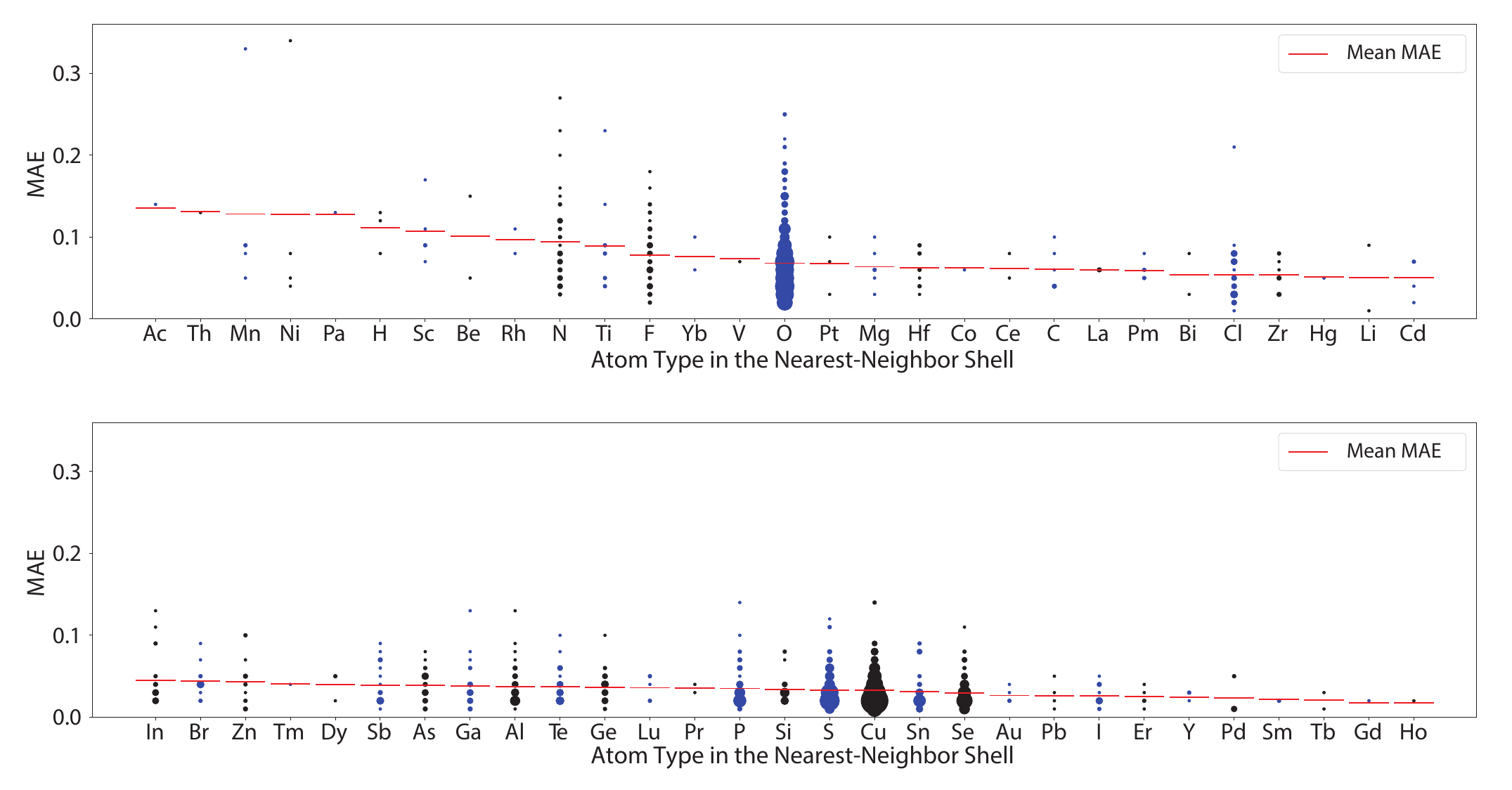}
    \caption{The accuracy of CuXASNet when predicting Cu spectra with different nearest neighbor elemental configurations. The x axis shows the nearest neighbors of the Cu absorber. Only predictions of the $L_3$ edge are visualized here. The spot size is proportional to the number of spectra at that point and the red horizontal lines denote the mean MAE for that category. A spectrum can be in more than one category if the Cu atom has multiple different elements in its nearest neighbors.}
    \label{fig:nearest_neighbor_analysis}
\end{figure*}

\begin{figure*}[ht]%
    \centering
    \includegraphics[width=\textwidth]{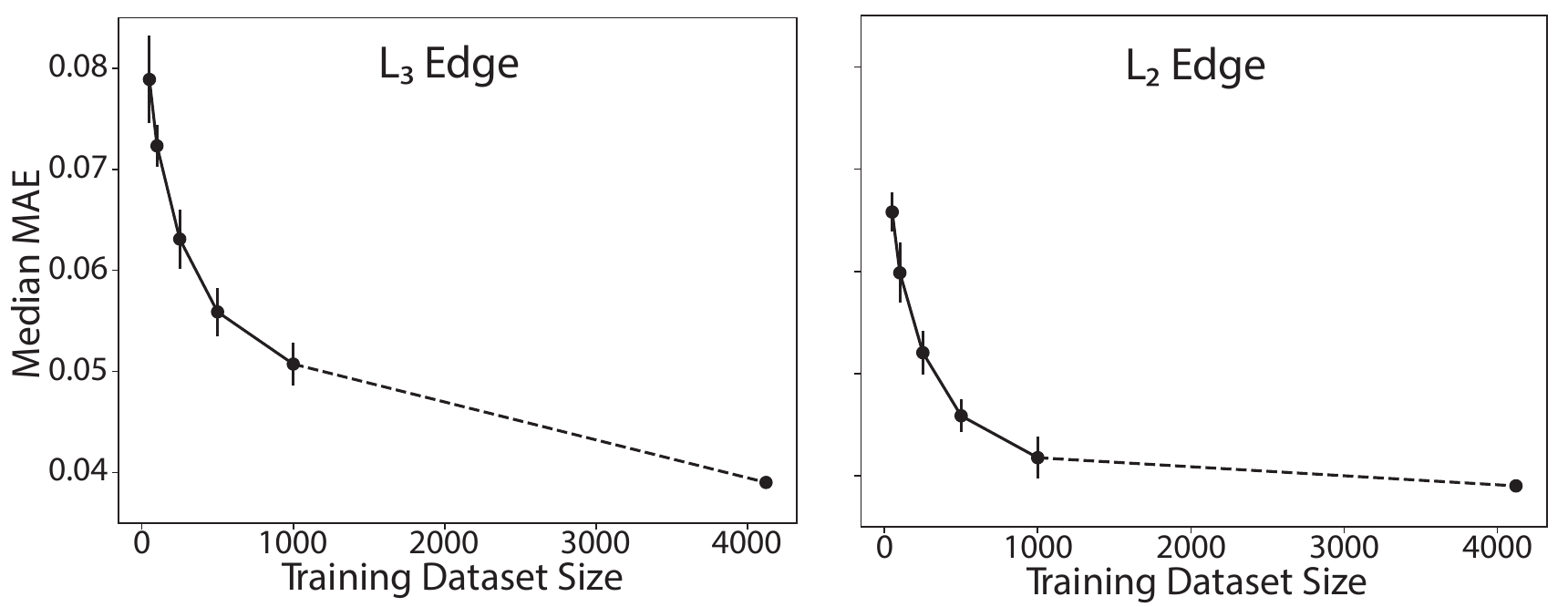}
    \caption{Accuracy of the model, as determined by the median MAE, when the model is applied to the test set using different volumes of training spectra. The smaller training sets are determined by random sampling the full training set used for the model presented in this work. The vertical bars on each scatter spot show two standard deviations of the median MAE when 10 different random samples of training data are used. In each case, the resulting model is predicted on the full test set to determine its accuracy.}
    \label{fig:training_set_size}
\end{figure*}

\subsection{Comparison to Experimental Spectra}
To determine the applicability of CuXASNet for analyzing experimental spectra, we validate its predictions against experimental spectra of the same materials. Figure~\ref{fig:experimental_comparison}(a-c) shows qualitative matching of spectra generated using CuXASNet (orange) to FEFF9 (blue) and experimental spectra (black). The three materials systems compared in the figure were determined by selecting the best, a middle, and the worst MAE between CuXASNet's predictions and experiments. By visual inspection, it is quite clear that the minor differences between the FEFF9 simulation and the ML predicted spectrum in Figure~\ref{fig:experimental_comparison}(a-b) often occur when CuXASNet produces a spectrum that is a slightly smoothed version of the FEFF9 spectra, and this causes no systematic difference when comparing to the experimental spectra. Even in the worst example shown in Figure~\ref{fig:experimental_comparison}(c), the error is found in the intensity of the $L_3$ edge rather than a misrepresentation of the features. This is confirmed quantitatively: Figure~\ref{fig:experimental_comparison}(d) demonstrates that the MAEs for FEFF9 simulation-experiments and CuXASNet-experiments are virtually identical across different materials, irrespective of CuXASNet-FEFF9 error. This confirms CuXASNet's potential to function as a rapid experimental analysis tool for determining the identity of an unknown experimental XAS spectrum. 

\subsection{Chemical Trends in Spectral Prediction Errors}
For the general use of this model, it is essential to understand the conditions under which it can be expected to perform well or struggle to reproduce FEFF9 simulations. Therefore, we explored the model performance versus OS and CN in Figure~\ref{fig:category_boxplot}. When analyzing the performance by OS, the trends reflect those seen in Figure~\ref{fig:dataset_visualization}, where the OS containing more spectral uniformity are also predicted with more accuracy on average. A general decrease in accuracy, as measured by the median MAE, is observed as OS increases. Although, the performance on Cu(0) and Cu(I) is very similar, and there are very few Cu(III) materials in the dataset, making their reliable prediction more challenging. An almost identical trend is observed in the $L_2$ edge, with the main difference being that there are fewer larger errors than observed in the $L_3$ edge prediction. The accuracy labeled by coordination environment is similarly reflected in Figure~\ref{fig:dataset_visualization}, with coordination environments that have a higher diversity of spectral features generally associated with less accurate spectral prediction. Interestingly, CuXASNet performs better on CNs in the ``other" category than CN=4, reflecting the wide range of core-level spectral features associated with 4-coordinated Cu atoms. 

\begin{figure*}[ht]%
    \centering
    \includegraphics[width=\textwidth]{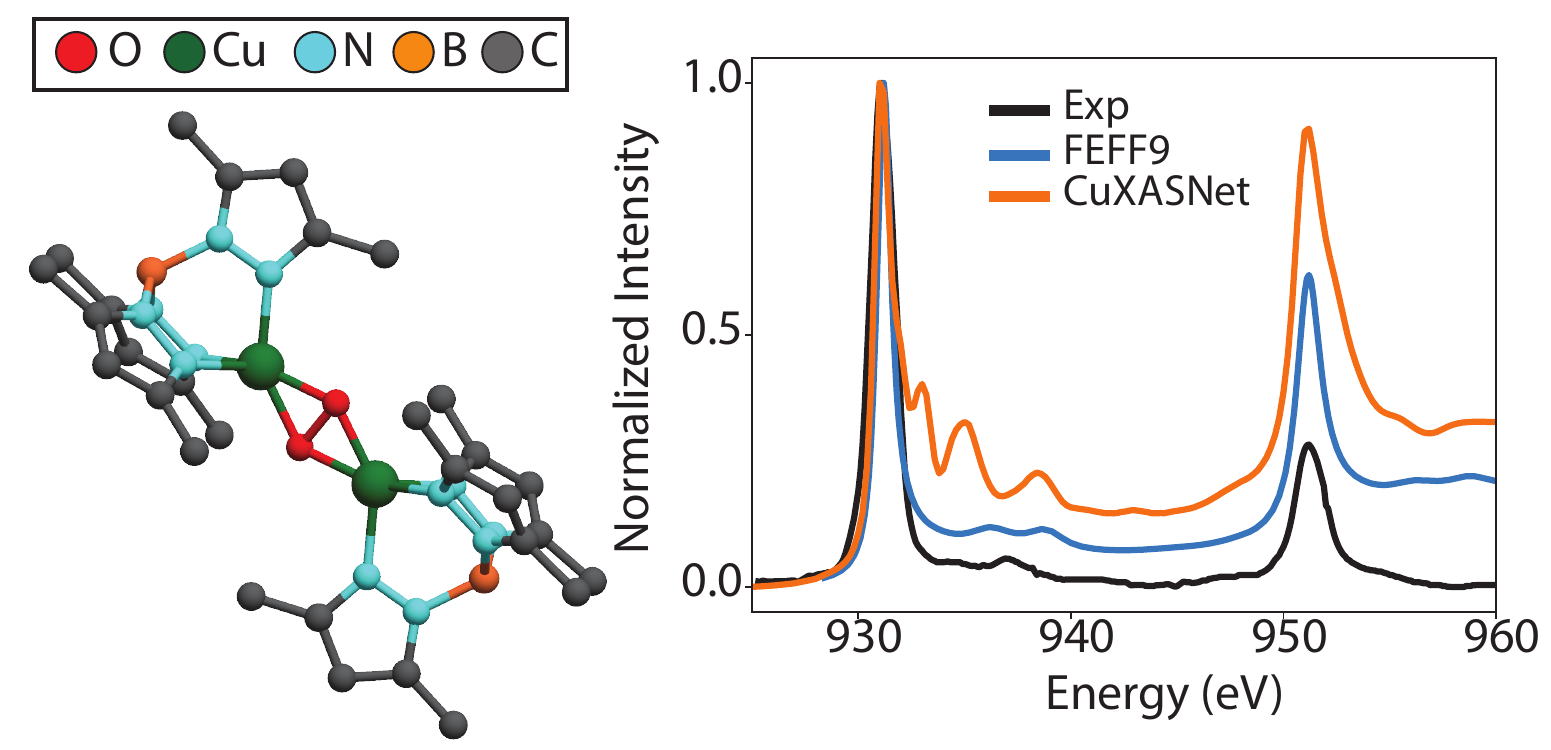}
    \caption{An application of our CuXASNet model to an uncommon and complicated Cu organometallic structure, extracted from Ref.~\onlinecite{Kitajima_1992_ANew}. A depiction of this structure is shown on the left, focusing on the region of the structure containing the Cu atoms. Hydrogen atoms are removed for clarity. A comparison between an experimental spectrum taken from Ref.~\onlinecite{Qayyum_2013_LEdge}, a FEFF9 simulation and the spectrum generated using CuXASNet is shown on the right. The spectra agree up to the baseline intensity between the $L_2$ and $L_3$ peaks.}
    \label{fig:complicated_Cu} 
\end{figure*}

CuXASNet's performance is broken down by the Cu absorber's nearest neighbors in Figure~\ref{fig:nearest_neighbor_analysis}. The two worst predictions from this model, which are significant outliers, contain Cu-Ni and Cu-Mn bonds. It is clear, however, that not all Cu-Ni and Cu-Mn bonds produce large errors. Although the sample size is quite low, there are 5 Cu absorbers that contain Cu-Mn bonds and  4 Cu absorbers that contain Cu-Ni bonds, and these two errors are significantly larger than the other examples of this type of bonding. A closer examination of these two structures shows they are antiperovskites of the formula A$_3$BX, where A is either Mn or Ni~\cite{Jain2013}. This is the only example of antiperovskite materials in the dataset, meaning users should be wary utilizing this model to predict absorbers in antiperovksite materials. Additionally, these two materials show a specific bonding structure of Cu-metal-N bonding, which may also be less accurate for CuXASNet. However, FEFF9 also generates a highly unusual spectrum for these materials, as shown in Figure~\ref{fig:worst_spec}. Therefore, this specific class of materials may be challenging for FEFF9 to model. Additionally, it does not appear to be the case that conventional Cu containing perovskite materials are worse than average for FEFF9 or this model. For example, the $L_3$ spectrum of the perovskite structure CsCu$_2$I$_3$ is predicted with a high accuracy, with an MAE of only 0.035. This corresponds to a log(MAE) of -1.45, which is slightly better than the median performing spectrum in Figure \ref{fig:sim_spec_accuracy}(c). 

Beyond the two outliers discussed above, general trends in performance by elemental composition show the worst errors among the well represented elements are commonly fluorides and nitrides. While many oxides are predicted quite accurately, there are a non-trivial amount of higher errors in Cu oxides as well. Beyond these, most Cu metallic alloys and Cu phosphides, sulfides and selenides are predicted accurately. Among the halogens, a clear positive correlation is seen between accuracy and atomic number, with F$<$Cl$<$Br$<$I, although iodides and bromides are not very well represented in the dataset.    

\subsection{Volume of Training Data Necessary For Inference}
One significant extension of this model is its potential applicability to more computationally intensive simulations, such as the OCEAN or VASP codes~\cite{Hafner_2008_Ab_initio,Vinson_2011_Bethe,Vinson_2022_Advances}. The model and analysis presented in this work was conducted using a dataset of 5497 spectra, with 4123 used for training. However, for systems like Ti and Fe, FEFF9 can struggle to represent the $L$-edge spectra, necessitating more computationally intensive simulations, the generation of which can be time and resource prohibitive. Therefore, in this work we have examined the volume of data necessary for accurate simulated spectral inference. This is visualized in Figure~\ref{fig:training_set_size}, which shows the model's accuracy on our test data as a function of training data volume. In this case, the smaller training set is random sampled from the 4000 spectra used for training the full CuXASNet model, with the test set held fixed. Figure~\ref{fig:training_set_size} shows that the increased performance from adding additional spectra begins diminishing at around 500 spectra, and that the $L_2$ edge reaches its steady state with fewer training spectra. This is unsurprising, given the relatively fewer significant spectral features in the $L_2$ edge (Figure~\ref{fig:dataset_visualization}) and the model's increased accuracy on the $L_2$ edge in general as shown in Figure~\ref{fig:sim_spec_accuracy}. The relatively low volume of data required to hit reasonable spectral prediction indicates that this model architecture could be used to generate spectra of higher computational cost without requiring a dataset that is computationally prohibitive. 

\subsection{Predictions Beyond Crystalline Structures}
To demonstrate the wide-ranging utility of CuXASNet on more atypical structures not originating from more standard crystalline materials, we have examined the literature for a Cu-containing molecule where an $L$-edge spectrum for Cu absorption had also been recorded~\cite{Qayyum_2013_LEdge}. The molecular complex examined here is a peroxo-Cu(II) species that is created when coupled bi-nuclear copper proteins bind molecular oxygen~\cite{Qayyum_2013_LEdge}. We extracted the structure of the complex from Ref.~\cite{Kitajima_1992_ANew} and compare the predicted Cu $L_{2,3}$ edge from CuXASNet with the experimental spectrum extracted from Ref.~\cite{Qayyum_2013_LEdge} in Figure~\ref{fig:complicated_Cu}. After manually aligning the $L_3$ peak energy to match the experimental spectrum, we observe reasonable agreement between CuXASNet and experiments up to the baseline intensity between the $L_3$ and $L_2$ peaks. The baseline intensity mismatch between the experimental spectrum and the FEFF9/CuXASNet spectra is an artifact of a processing procedure done in Ref.~\onlinecite{Qayyum_2013_LEdge} in order to conduct density of state calculations. Despite this, CuXASNet overestimates the $L_2$ edge intensity relative to the $L_3$ edge when comparing to the FEFF9 simulation for this complex as well. Additionally, the two small peaks after the $L_3$ edge predicted by FEFF9 are generally mispredicted by CuXASNet, and CuXASNet also predict a spurious shoulder after the $L_3$ edge. However, despite these minor mispredictions, the overall shape of the $L_3$ and $L_2$ edges are rendered reasonably well. This example shows a success case for CuXASNet on a structure far outside the realm of its training data and demonstrates that CuXASNet is a viable tool for fast spectral screening of candidate structures for an unknown system, regardless of structural complexity. This can inform structural candidates for manual simulation studies and help verify an unknown material's composition and structural parameters. To further enhance CuXASNet's ability to generate accurate spectra of unknown complicated structures, CuXASNet can be dynamically refined using an active learning loop.

\section{Conclusion}
This work showcases CuXASNet, an ML model which can predict the $L$-edge XAS spectrum of Cu based on the material's atomic structure. CuXASNet utilizes a graph neural network, M3GNet, to featurize materials or clusters into a fixed length vector representation, which is then fed to the neural network for spectral prediction. CuXASNet predicts the $L_3$ and $L_2$ edges for Cu sites with quantitative accuracy to FEFF9. The model shows excellent accuracy on Cu alloys and is highly reliable on most oxide materials. Additionally, our predicted spectra have good agreement to experimental spectra from the literature. Our generated spectra match experiments with an average Spearman's correlation coefficient of 0.891 and an average MAE of 0.125,  matching FEFF9's experimental comparison values of 0.898 and 0.131 for these metrics. We also demonstrate that CuXASNet is able to predict spectra for structures outside of conventional crystalline materials by turning it to a complicated organometallic Cu molecule. The accuracy of CuXASNet on simulated data, and the viability of the experimental comparison relative to simulated spectra, shows CuXASNet is a highly valuable tool for rapidly generating more experimentally viable simulated data. There are many interesting potential applications for this model, including structural determination from spectra, targeted materials synthesis, and the rapid and accurate generation of training data for ML models. Additionally, we introduce a model framework that can be extended to generate spectra of more computationally intensive simulations and other transition metal $L$-edges.

\section{Data and Code Availability}
The spectral dataset and the code to generate and analyze CuXASNet can be found in the GitHub repository https://github.com/smglsn12/CuXASNet. Due to the unpublished nature of this work, this repository is currently private, but will be shared upon request. This repository will be made public upon publication of this work.  

\section{Author Contributions} 
SPG generated the simulated XAS dataset, conducted the machine learning training and analysis, and wrote the manuscript. MRC developed the automated procedure to generate input files for FEFF9 spectral simulation, provided access to the M3GNet model used in this work, and provided advising on the development of the machine learning models. DL provided training and expertise necessary to generate the simulated dataset and provided project direction and advising. JC provided experimental core-level spectroscopy knowledge, led the collaboration and designed the scope of this work. All authors read, edited and approved the final manuscript.

\section{Acknowledgment}
This work was primarily funded by the US Department of Energy in the program “4D Camera Distillery: From Massive Electron Microscopy Scattering Data to Useful Information with AI/ML.” Work at the Molecular Foundry was supported by the Office of Science, Office of Basic Energy Sciences, of the U.S. Department of Energy under Contract No. DE-AC02-05CH11231. This research is supported by the U.S. Department of Energy,
Office of Science, Office Basic Energy Sciences, Award Number FWP
PS-030, and used Theory and Computation resources of the Center for Functional Nanomaterials (CFN), which is a U.S. Department of Energy Office of Science User Facility, at Brookhaven National Laboratory under Contract No. DE-SC0012704. 

\bibliography{references_manual}   

\setcounter{figure}{0}
\renewcommand{\figurename}{Fig.}
\renewcommand{\thefigure}{S\arabic{figure}}

\begin{figure*}[ht]%
    \centering
    \includegraphics[width=\textwidth]{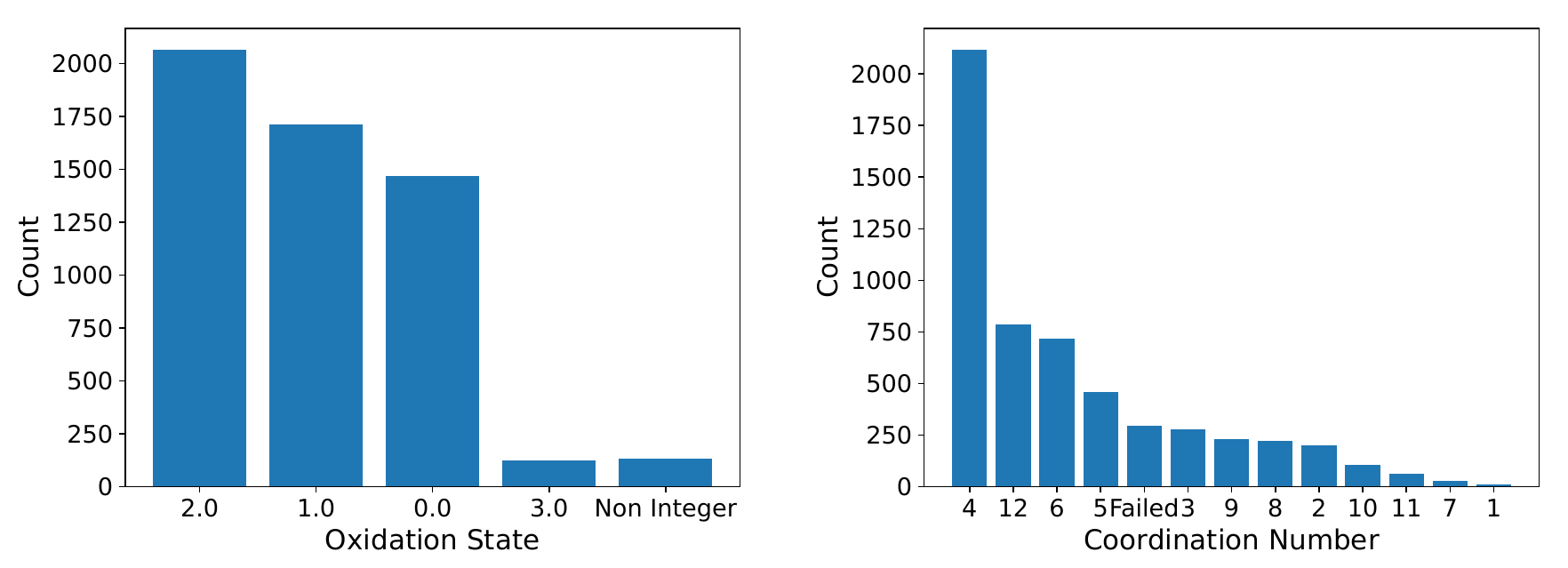}
    \caption{Bar plots showing the number of spectra corresponding to each OS (left) and CN (right). In the CN plot, values other than 4, 12 and 6 are represented by the 'other' category shown in Figure~\ref{fig:dataset_visualization}. The 'Failed' marker denotes the Pymatgen function used to determine the CN failed to report a value \cite{Jain2013}.}
    \label{fig:data_distribution} 
\end{figure*}

\begin{figure*}[ht]%
    \centering
    \includegraphics[width=\textwidth]{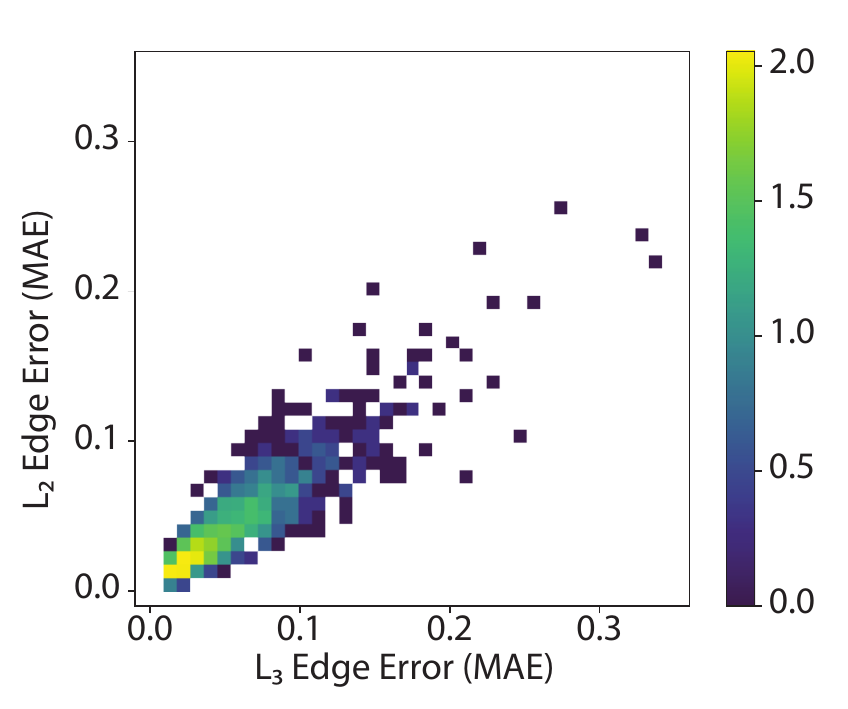}
    \caption{The relationship between the $L_3$ MAE and the $L_2$ MAE for the same absorbing site. The color denotes the number of spectra at that point, which is transcribed onto a log scale for ease of viewing outliers.}
    \label{fig:edge_correlation} 
\end{figure*}

\begin{figure*}[ht]%
    \centering
    \includegraphics[width=16cm]{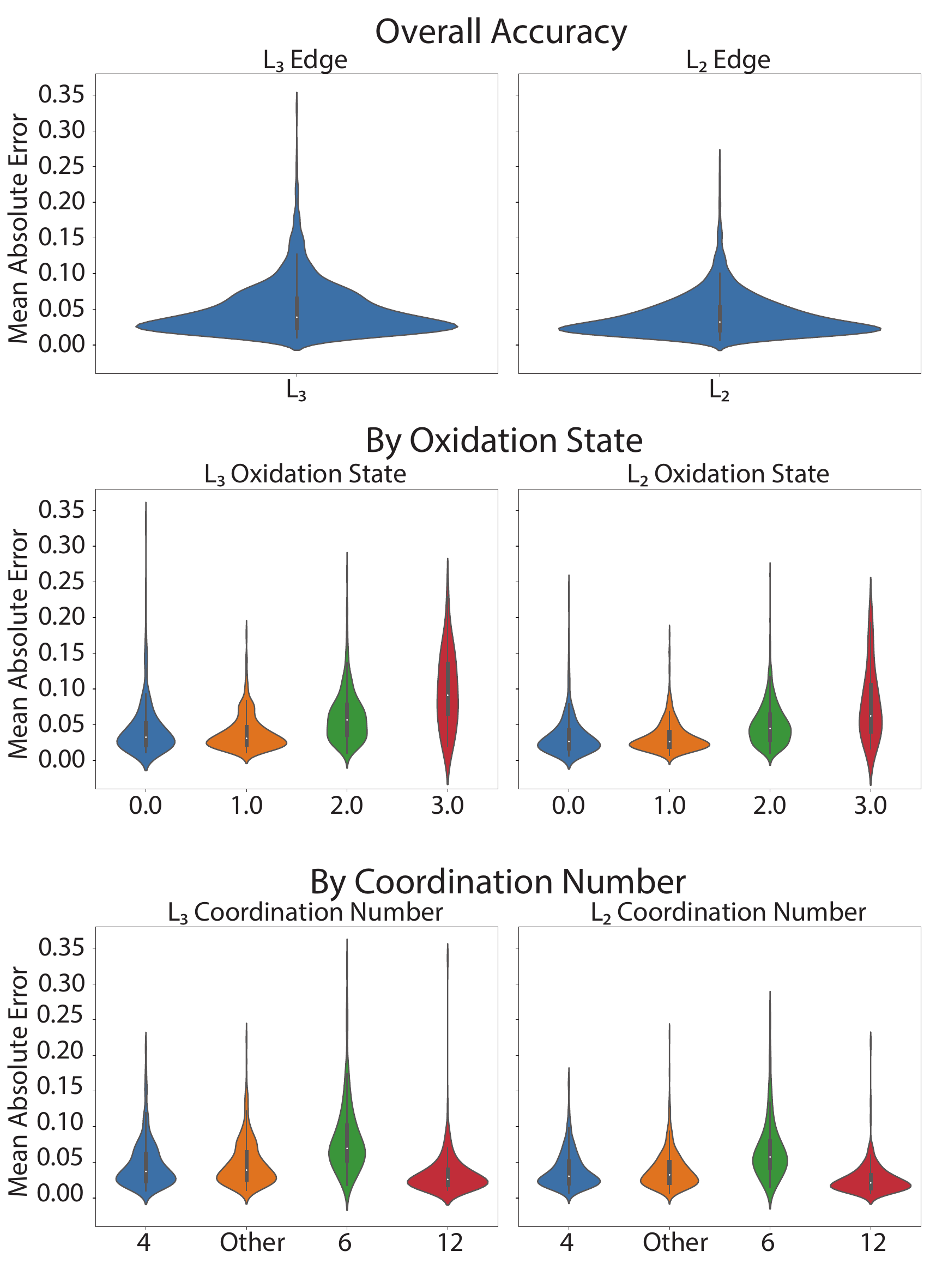}
    \caption{A detailed visualization of the spread of model accuracy by various chemical labels. Each violin plot indicates the data distribution, where the thick gray bar inside the violins indicates the interquartile range.}
    \label{fig:category_boxplot} 
\end{figure*}

\begin{figure*}[ht]%
    \centering
    \includegraphics[width=\textwidth]{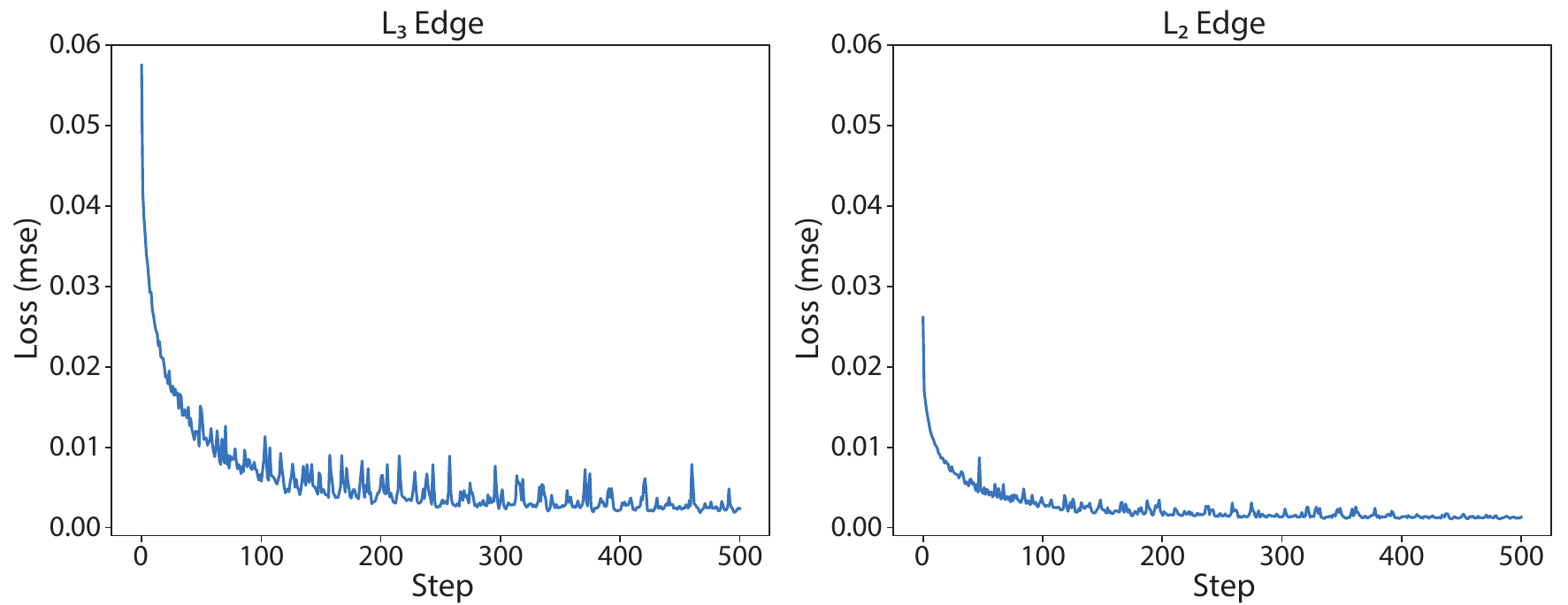}
    \caption{The training loss, measured by MSE of the predicted spectrum vs the true spectrum, vs model training step. }
    \label{fig:training_loss} 
\end{figure*}

\begin{figure*}[ht]%
    \centering
    \includegraphics[width=\textwidth]{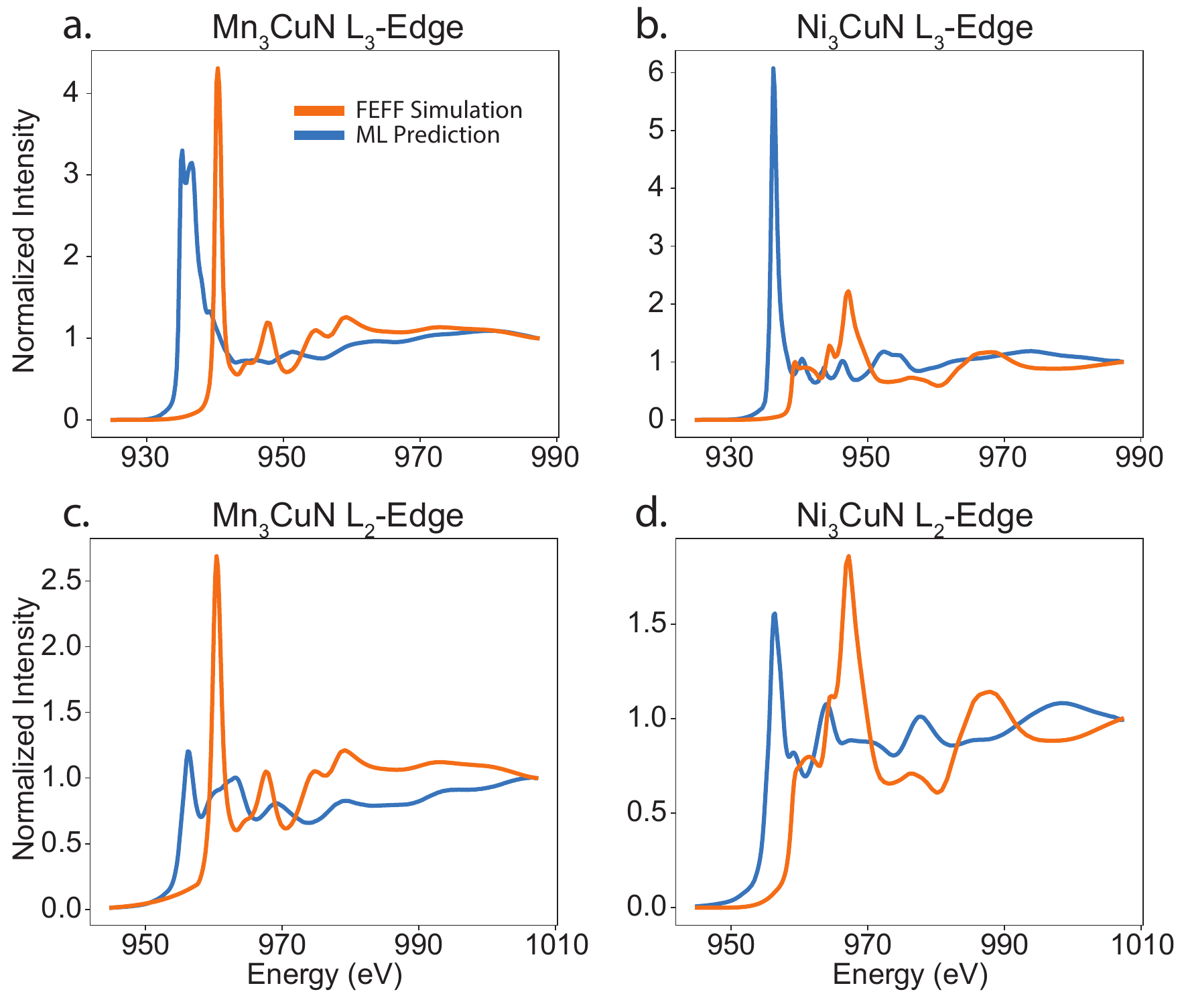}
    \caption{Depictions of the two least accurate predictions of CuXASNet. The $L_3$ edge, which is far less accurate, is shown in (a) and (b) for Mn$_3$CuN and Ni$_3$CuN, respectively. The $L_2$ edge is shown in (c) and (d) for Mn$_3$CuN and Ni$_3$CuN, respectively. These samples are characterized by FEFF9 simulations (shown in orange) that differ significantly from the spectra shown in Figure~\ref{fig:dataset_visualization}, especially in the energy location of the edge onset, which is roughly 10 eV higher than most other Cu $L$-edge spectra.}
    \label{fig:worst_spec} 
\end{figure*}

\end{document}